\newif\ifarxiv
\arxivtrue 

\ifarxiv
\documentclass{jfm-arxiv}
\else
\documentclass[lineno]{jfm}
\fi
\usepackage{graphicx}
\usepackage{newtxtext}
\usepackage{newtxmath}
\usepackage{natbib}
\usepackage{hyperref}
\usepackage{xcolor}
\usepackage{multirow}
\hypersetup{
    colorlinks = true,
    linkcolor  = blue,
    urlcolor   = blue,
    citecolor  = blue,
}

\newcommand{\RomanNumeralCaps}[1]

\def\drawline#1#2{\raise 2.5pt\vbox{\hrule width #1pt height #2pt}}

\usepackage{soul}

\def\mytitle{On the relevance of lift force modelling in turbulent wall flows with small inertial particles}
\def\myshorttitle{Relevance of lift in particulate turbulent wall flows}
\ifarxiv
\title[\myshorttitle]{\vspace{-85pt}\mytitle}
\else
\title[\myshorttitle]{\mytitle}
\fi

\author{Wei Gao\aff{1,2}
  \corresp{\email{wei.gao@kaust.edu.sa}},
  Pengyu Shi\aff{3,4},
  Matteo Parsani\aff{1,2} \and
  Pedro Costa\aff{5}}

\shortauthor{W.~Gao, P.~Shi, M.~Parsani and P.~Costa}

\affiliation{\aff{1}Mechanical Engineering, Physical Science and Engineering Division, King Abdullah University of Science and Technology, Thuwal 23955-6900, Saudi Arabia
\aff{2}Applied Mathematics and Computational Science, Computer Electrical and Mathematical Science and Engineering Division, Extreme Computing Research Center, King Abdullah University of Science and Technology, Thuwal 23955-6900, Saudi Arabia
\aff{3}Institut de M\'ecanique des Fluides de Toulouse (IMFT), Universit\'e de Toulouse, CNRS, Toulouse, France
\aff{4}Helmholtz-Zentrum Dresden – Rossendorf, Institute of Fluid Dynamics, 01328 Dresden, Germany
\aff{5}Process \& Energy Department, TU Delft, Leeghwaterstraat 39, 2628~CB Delft, The Netherlands}

\begin{document}
\maketitle
\ifarxiv
\thispagestyle{empty}
\fi

\begin{abstract}
In particle-laden turbulent wall flows, lift forces can influence the near-wall turbulence. This has been recently observed in particle-resolved simulations, which, however, are too expensive to be used in upscaled models. Instead, point-particle simulations have been the method of choice to simulate the dynamics of these flows during the last decades. While this approach is simpler, cheaper, and physically sound for small inertial particles in turbulence, some issues remain. In the present work, we address challenges associated with lift force modelling in turbulent wall flows and the impact of lift forces in the near-wall flow. We performed direct numerical simulations (DNS) of small inertial point particles in turbulent channel flow for fixed Stokes number and mass loading while varying the particle size. Our results show that the particle dynamics in the buffer region, causing the apparent particle-to-fluid slip velocity to vanish, raises major challenges for accurately modelling lift forces. While our results confirm that lift forces have little influence on particle dynamics for sufficiently small particle sizes, for inner-scaled diameters of order one and beyond, lift forces become quite important near the wall. The different particle dynamics under lift forces results in the modulation of streamwise momentum transport in the near-wall region. We analyze this lift-induced turbulence modulation for different lift force models, and the results indicate that realistic models are critical for particle-modelled simulations to correctly predict turbulence modulation by particles in the near-wall region.
\end{abstract}



\section{Introduction}
\label{sec:headings}

Wall-bounded turbulent flows laden with inertial particles abound in environmental and industrial contexts, such as the transport of particulate matter in the atmosphere, sediment transport in rivers, the separation of fine particles within industrial cyclones, and fluidized bed reactors. The chaotic and multiscale nature of the fluid turbulence coupled with the particle dynamics results in fascinating phenomena, which, however, are still challenging to understand and model \citep{balachandar2010turbulent,brandt2022particle}.

In the past decades, many studies have been devoted to the dynamics of turbulent wall transport of small inertial particles, driven by their prevalent nature and rich physics. These systems have been traditionally classified by how the dispersed phase influences the overall flow behaviour \citep{Elghobashi-ASR-1994}. Precisely, one-way coupling (1WC) corresponds to very small particle loadings, so small that their influence on the dynamics of the suspending fluid turbulence is negligible; two-way coupling (2WC) denotes flow regimes where mass loading is high enough such that the particles modify flow observably, but particle-particle interactions are negligible; finally regimes where both particle--particle and particle--fluid interactions influence the flow dynamics are grouped into the four-way coupling category.

In a first-principles, fully-resolved direct numerical simulation (DNS) of a particle-laden flow, the flow around each particle needs to be explicitly resolved \citep{balachandar2010turbulent,Maxey-ARFM-2017} (so-called particle-resolved DNS, PR-DNS). While this approach is free from modelling assumptions for the dispersed phase dynamics, it is computationally expensive due to the explicit imposition of no-slip and no-penetration boundary conditions at the surface of many particles moving in a turbulent medium. This is incredibly challenging when particles are tiny and there is a scale separation between the particle size and the smallest (Kolmogorov) turbulence scale due to the need for extreme resolution requirements. Fortunately, in this case, one may be able to resort to the so-called point-particle approximation (PP-DNS), where interphase coupling is considered to be localized to a point. 

Indeed, the point-particle approximation has been the method of choice for simulating the particle dynamics in turbulent wall flows. In these cases, it is assumed that the local properties of an undisturbed flow at the particle position drive the dispersed-phase dynamics \citep{Maxey-Riley-PoF-1983,Gatignol-1983}. In the case of highly inertial particles (i.e., large particle-to-fluid density ratios), the so-called Maxey-Riley-Gatignol equations simplify to a drag force term \citep{arcen2006influence}, which nevertheless yields highly non-trivial particle dynamics, even in isotropic turbulence \citep{Toschi-and-Bodenschatz-ARFM-2009}. In wall-bounded turbulent flows, the inhomogeneous turbulence results in even richer dynamics, with the particle distribution driven by the interplay between small-scale clustering, turbophoresis, and the interaction between the particles and near-wall turbulence structures \citep{reeks1983transport,soldati2009physics,sardina2012wall}, resulting in very inhomogeneous particle concentrations peaking at the wall, with strong preferential sampling of low-speed regions, as reproduced in a plethora of numerical studies, such as \cite{Fessler-et-al-PoF-1994,Uijttewaal-and-Oliemans-PoF-1996,Marchioli-et-al-IJMF-2003,Kuerten-PoF-2006,Marchioli-et-al-IJMF-2008,soldati2009physics,Bernardini-et-al-IJMF-2013,jie2022existence}.

When two-way coupling effects are important, a point-particle DNS must describe the back-reaction of the dispersed phase in the flow. This is a major challenge, as the point-particle dynamics are driven by a local \emph{undisturbed} fluid velocity, while the local flow field is being disturbed by the particles \citep[see, e.g.][]{Gualtieri-et-al-JFM-2015}. The classical approach, known as the particle-in-cell method, was developed by \citet{Crowe-et-al-JFE-1977} and is widely used -- even in the present study -- but it requires sufficiently high number of particles per grid cell and cannot reproduce simple benchmarks of a sedimenting sphere in a quiescent medium. Indeed, approaches for a consistent treatment are being actively investigated; see \cite{Gualtieri-et-al-JFM-2015,Horwitz-and-Mani-JCP-2016,Ireland-and-Desjardins-JCP-2017,horwitz2020two,horwitz2022discrete}. Investigations of  particle-laden turbulent flows in the two-way-coupling regime are found in e.g.\ \cite{Vreman-et-al-FTC-2009,zhao2010turbulence,Kuerten2016point,richter2014,capecelatro2018transition,wang2019two,battista2023drag}.

While employing the point-particle approximation to small inertial particles is physically sound, validating the fidelity of the approximation in one- and two-way coupling conditions remains a challenge. Experimental data are available \citep{Eaton-and-Fessler-IJMF-1994,Kaftori-and-Banerjee-PoF-1995}. Still there are few parameter-matched numerical studies due to limitations in terms of Reynolds number and well-controlled experiments, and only in recent years, efforts in this direction have started to appear \citep{wang2019inertial}. Fortunately, PR-DNS of small inertial particles in turbulence has become possible, thanks to the continuous growth in available computer power and development of efficient numerical methods, with the first direct comparisons between point-particle models and particle-resolved simulations starting to appear for forced homogeneous isotropic turbulence and decaying HIT with moving particles \citep{Schneiders-et-al-F-2017,Schneiders-et-al-JFM-2017,Mehrabadi-et-al-JFM-2018,Frohlich-et-al-FTC-2018}, and turbulent channel flow \citep{Horne-and-Manesh-JCP-2019,costa2020interface}.

These results from PR-DNS of particles in wall-bounded turbulence confirm that a sole drag force may not suffice to accurately reproduce the particle dynamics even for relatively small particles with a large density ratio under one-way coupling conditions. Where the shear rate is high near the wall, \emph{lift} forces are also important. This has already been suggested in early works using PP-DNS, since the work by \cite{mclaughlin1989aerosol} \cite[see also][]{botto2005effect}. In this regard, while \cite{arcen2006influence} reported that the lift force has a negligible impact on the dispersed phase statistics, \cite{marchioli2007influence} and \cite{shin2022dynamics} found that the inclusion of the lift force in PP-DNS can lead to weakened near-wall particle accumulation in upward and horizontal channels, respectively; \cite{mclaughlin1989aerosol} showed that the inclusion of the lift force resulted in a higher deposition rate; \cite{wang1997role} found that neglecting the lift force resulted in a slight reduction in the deposition rate.

While these findings are not necessarily contradictory, as there are some variations in the governing parameters in the different studies and lift force models, many questions remain elusive: \emph{(1) which form of lift force model is appropriate for reproducing with high fidelity the dynamics of small inertial particles in a turbulent wall flow; (2) under which conditions do lift forces matter in the particle dynamics; and (3) what are the consequences of choosing an inaccurate lift force model in dispersed phase dynamics and near-wall turbulence modulation.}

In the recent direct comparison between PR-DNS and PP-DNS for small particles in turbulent channel flow by \cite{costa2020interface} \cite[see also][]{costa2020corr}, it was shown that lift models are vital for reproducing the near-wall particle dynamics, at least for inertial particles with a size of the order of one viscous wall unit. Surprisingly, a modified Saffman lift model predicted the particle dynamics perfectly, with the Saffman force scaled by a normalized shear rate, while conventional lift models yield poorer predictions \citep{costa2020corr}. In addition to reducing the near-wall concentration peak, it was also seen that lift force causes a large increase in correlated streamwise--wall-normal particle velocity fluctuations. This quantity has dramatic drag-increasing effects for sufficiently high mass loading, as it modulates the streamwise momentum budget; see \cite{costa2021near}. This direct link between lift force and drag increase makes their accurate modelling crucial.  

In the present work, we address the three questions presented above by performing DNS of turbulent channel flow laden with small inertial particles, using the point-particle approximation, with three different lift models from the classical Saffman lift model to the one that perfectly predicts previous PR-DNS data. We consider different inner-scaled particle sizes $D^+\sim 1-0.1$ in one and two-way coupling conditions for a Stokes number that is known to feature strong wall accumulation and preferential concentration in low-speed regions. Our analysis shows that currently available lift models are bound to fail near the wall for small inertial particles due to a vanishing particle-to-fluid slip velocity. We then use two-way coupling simulations to illustrate how different lift force models can result in \emph{qualitatively} different turbulence modulation (i.e., turbulence attenuation vs. turbulence enhancement). Near-wall accumulation is still significantly reduced for the smallest particle size ($D^+=0.1$). Still, the lift-induced increase of correlated velocity fluctuations near the wall is negligible, and thus, lift force has little consequences in turbulence modulation.

This paper is organized as follows. Next, in \S\ref{sec:numerical}, the numerical method, lift force models, and computational setups are described. Then, in \S\ref{sec:res}, the effects of lift force models on the particle dynamics and near-wall accumulation are investigated using one-way coupling DNS, based on which we try to explore the reason for the failure of conventional lift force models. Following this, we qualitatively demonstrate the effect of lift force models on turbulence modulation and momentum transfer with two-way coupling DNS. Finally, conclusions are drawn in \S\ref{sec:conclusion}.

\section{Methodology}
\label{sec:numerical}

\subsection{Governing equations and numerical method}\label{sec:model}
The fluid phase is governed by the incompressible Navier--Stokes equations,
\begin{equation}
\begin{gathered}
\nabla\cdot\mathbf{u}=0, \\
\rho\left(\partial_t \mathbf{u} + (\mathbf{u}\cdot\nabla)\mathbf{u}\right)=-\nabla p + \mu\nabla^2\mathbf{u} + \mathbf{f},
\end{gathered}
\label{ge_fluid}
\end{equation}
where $\mathbf{u}$ denotes the fluid velocity vector, $p$ is the fluid pressure, $\mu$ is the dynamic viscosity, $\rho$ is the fluid density, and $\mathbf{f}$ is the particle feedback force to the fluid phase in the case of two-way coupling point-particle simulations, here computed by using a standard particle-in-cell approach, which spreads the particle force to the nearest Eulerian grid points with a linear kernel \citep{boivin1998direct,lee2019effect,zhang2023electrostatic}.

We consider $x$, $y$, and $z$ as the streamwise, wall-normal, and spanwise directions, respectively. No-slip and no-penetration boundary conditions are specified at the domain walls, while periodicity is applied along the streamwise and spanwise directions. These equations are discretized in space using a pseudospectral approach along $x$ and $z$ and second-order finite differences along $y$. Wray's low-storage third-order Runge--Kutta scheme is employed for time marching \cite[see][]{wray1990}.

Spherical particles with density $\rho_p \gg \rho$ and particle diameter $D$ in the absence of gravity are tracked with a standard Lagrangian point-particle method, with their dynamics governed by
\begin{equation}
m_p \dot{\mathbf{U}}_p = \mathbf{F}_d+\mathbf{F}_l\mathrm{,} \;\; \dot{\mathbf{X}}_p = \mathbf{U}_p\mathrm{,}
\end{equation}
with the Schiller--Naumann drag force
\begin{equation}
\mathbf{F}_d = -3\pi\mu D \mathbf{U}_s\left(1+0.15\Rey_p^{0.687}\right)\mathrm{,}
\label{eqn:drag}
\end{equation}
where $\mathbf{U}_p$, $\mathbf{X}_p$, $\mathbf{F}_d$ and $\mathbf{F}_l$ are the particle velocity, position, drag and lift forces, $m_p$ is the particle mass, $\mathbf{U}_s=\mathbf{U}_p-\mathbf{u}|_{\mathbf{x}=\mathbf{X}_p}$ is the local slip velocity evaluated at the particle location, and $\Rey_p$ $=\left|\mathbf{U}_s\right| D / {\nu}$ is the particle Reynolds number. Since the particle-to-fluid density ratio is high, other dynamic effects such as added mass, fluid acceleration, and the Basset history force are negligible. While in most practical scenarios of particle-laden wall transport, gravity is important at high density ratios, we neglect it in the present work. We do this to isolate the interplay between more intricate wall accumulation mechanisms (e.g., turbophoresis) and lift forces, in a flow which could otherwise feature significant settling effects. Finally, for the sake of simplicity, a perfectly elastic hard-sphere rebound is employed for particle--wall interactions.

\subsubsection*{Lift force models}

We consider standard shear-induced lift force models where lift force $\mathbf{F}_l$ acts perpendicular to the local shear sampled by the particle and points along $\boldsymbol{\omega}\times\mathbf{U}_s$, where $\boldsymbol{\omega} = \left(\nabla \times \mathbf{u} \right)|_{\mathbf{x}=\mathbf{X}_p}$ is the undisturbed flow vorticity evaluated at the particle position. For convenience, let us express the lift force in terms of the dimensionless lift coefficient, $C_L$, defined as
\begin{equation}
C_L = \frac{\mathbf{F}_l}{\frac{1}{8}\pi\rho D^2 |\mathbf{U}_s|^2}\cdot\frac{\boldsymbol{\omega}\times\mathbf{U}_s}{|\boldsymbol{\omega}\times\mathbf{U}_s|}.
\label{eq:cl}
\end{equation}
Assuming the Oseen length $\ell_u=\nu/{\left| \mathbf{U}_s \right|}$ is much larger than the Saffman length $\ell_\omega=\sqrt{\nu/{\left| \boldsymbol{\omega} \right|}}$, i.e. $\varepsilon \gg 1$ with $\varepsilon=\sqrt{\left| \boldsymbol{\omega}\right| \nu} /\left| \mathbf{U}_s \right|$, an explicit lift solution can be derived:
    \begin{equation}
    C_L = \frac{18}{\pi^2}\varepsilon J(\varepsilon),
        \label{eq:cl-base}
    \end{equation}
    with
    \begin{equation}
    J(\varepsilon\gg1) =J^\infty=2.255.
        \label{eq:cl-sa}
    \end{equation}
Hereafter, the lift expression~\eqref{eq:cl-base} together with the $J$ function by Eq.~\eqref{eq:cl-sa} will be referred to as the Saffman model, which is expected to be valid in the double limits $\Rey_p\to 0$ and $\Rey_\omega\to 0$ provided that $\varepsilon \gg1$, where $\Rey_\omega=\left| \boldsymbol{\omega} \right|D^2/\nu$ is the shear Reynolds number.

Still in the double limits $\Rey_p\to 0$ and $\Rey_\omega\to 0$, the $J$ function in Eq.~\eqref{eq:cl-base} at finite $\varepsilon$ turns out to be a volume integral in Fourier space. Its value cannot be put in closed form but was estimated numerically in \cite{asmolov1989lift}, \cite{mclaughlin1991inertial}, and more recently, \cite{shi2019lift}. Based on these numerical data, various empirical correlations of $J(\varepsilon)$ were proposed (see \cite{shi2019lift} for a comprehensive review), with the most commonly used being the one proposed by \cite{mei1992approximate}, i.e.,
\begin{equation}
J(\varepsilon) =0.3J^{\infty} \left(1+\tanh \left[\frac{5}{2}\left(\log _{10} \varepsilon+0.191\right)\right]\right)\left(\frac{2}{3}+\tanh (6 \varepsilon-1.92)\right).
\label{eq:cl-mei}
\end{equation}
Hereafter, the lift expression~\eqref{eq:cl-base} together with the $J$ function by Eq.~\eqref{eq:cl-mei} will be referred to as the Mei model, which is expected to be valid irrespective of $\varepsilon$, provided that $\Rey_p$ and $\Rey_\omega$ are small.

The flow is assumed to be unbounded for the two lift models introduced above. In particle-laden channel flows, near-wall accumulation of particles is often observed, as discussed above. Typically, the peak in the particle concentration appears within the viscous sublayer, i.e., for $y^+ \leq 5$, wherein the wall effect on the lift is crucial if $D^+ \geq 1$ \citep{balachandar2010turbulent,shi2021drag}. In this context, the explicit lift solution can be derived if the wall lies in the ``inner region'' of the flow disturbance in the low-$\Rey_p$ limit, specifically, if the separation between the particle and the wall, $\ell$, is much smaller than the inertial lengths ($\ell \ll \min{(\ell_u, \ell_\omega)}$). The corresponding lift solutions (see \cite{shi2020lift} for a comprehensive review) take the general form
\begin{equation}
C_L = A + B Sr + C Sr^2,
\label{eq:cl-in}
\end{equation}
where $Sr= \left| \boldsymbol{\omega} \right| D/\left| \mathbf{U}_s \right|$ is the normalized shear rate, and $A$, $B$, and $C$ are pre-factors that are independent of $Sr$. Hereafter, the terms proportional to $Sr$ and $Sr^2$ will be referred to as the linear and quadratic inner contributions $C_{L, \text{in}}^\text{linear}$ and $C_{L, \text{in}}^\text{quad}$, respectively.

Situations where the wall is in the ``outer region'' of the flow disturbance ($\ell \gg \max{(\ell_u, \ell_\omega)}$), while still in the limit $\Rey_p\to 0$, were considered in \cite{asmolov1990dynamics}, \cite{mclaughlin1993lift} and \cite{takemura2009migration}. Assuming that $Sr=\Rey_\omega^{1/2}\varepsilon\leq O(1)$ (i.e. the same condition for the shear rate as considered in \citet{saffman1965lift}), the outer-region lift approaches the Saffman solution~\eqref{eq:cl-base}, namely $C_{L, \text{out}}^\text{linear}\propto \varepsilon$. Together with the scaling in the inner-region that $C_{L, \text{in}}^\text{linear}\propto Sr$, it appears that the inner-region lift contribution transitions into the outer-region by a pre-factor $\varepsilon /Sr$. If one assumes that the quadratic contribution follows the same transition, it may be speculated that $C_{L, \text{in}}^\text{quad}$ scales as $\varepsilon Sr$ in the outer region, i.e. $C_{L, \text{out}}^\text{quad} \propto \varepsilon Sr$. For the particle-laden channel flow,
\begin{equation}
Sr\approx\frac{(u_\tau^2/\nu)D}{u_s}=\left(\frac{u_s}{u_\tau}\right)^{-1}D^+,
\label{eq:Sr}
\end{equation}
where $u_s$ is the streamwise mean slip velocity, namely $u_s=\langle\mathbf{U}_s\cdot\mathbf{e}_x\rangle$ with $\mathbf{e}_x$ the unit vector along the streamwise direction; $u_\tau$ is the conventional wall friction velocity. As will be demonstrated in figure \ref{fig:up_1wc} in section \ref{sec:1wc}, $u_s$ might change its sign in the buffer layer, leading to extremely large values of $Sr$ in the near-wall region.

The analysis above implies that there is also a quadratic lift contribution $C_{L}^\text{quad} \propto \varepsilon Sr$, which might dominate the lift generation in the inner wall region. This is consistent with the form of the lift force proposed in \cite{costa2020interface,costa2020corr}, where
\begin{equation}
C_L = \frac{18}{\pi^2}\varepsilon Sr J^\infty,
\label{eq:cl-co}
\end{equation}
with $J^\infty=2.255$ according to \cite{saffman1965lift}. Despite its simplified form, this correlation aligns well with their PR-DNS results. Hereafter, the lift model by Eq.~\eqref{eq:cl-co} will be referred to as the CBP model.

Finally, it should be noted that in practice, to avoid singularities in the numerical calculation, we implemented a slight variant of the lift coefficient as described in \eqref{eq:cl}, where the quotient on the right-hand side to define the unit vector is modified to $\boldsymbol{\omega}\times\mathbf{U}_s/|\boldsymbol{\omega}||\mathbf{U}_s|$. In practice, the results are not very sensitive to this choice, as $\boldsymbol{\omega}\times\mathbf{U}_s$ is nearly aligned with the wall-normal direction.

\subsection{Validation}
\begin{figure}
\centerline{\includegraphics[width=1.0\textwidth,keepaspectratio]{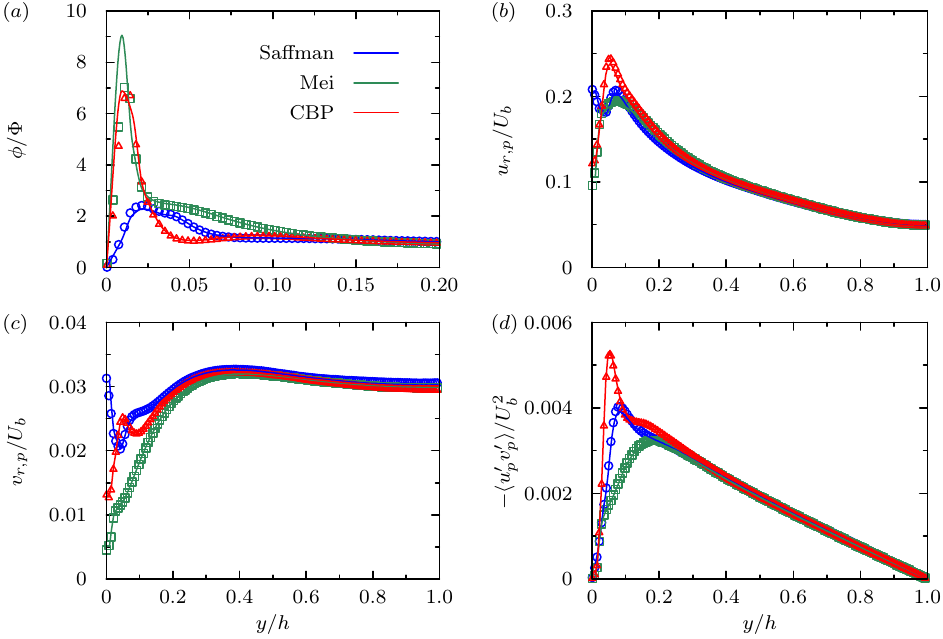}}
\caption{$(a)$ Normalized particle volume fraction $\phi/\Phi$ and outer-scaled second-order moments of particle velocity: $(b)$ streamwise velocity r.m.s., $(c)$ wall-normal velocity r.m.s., $(d)$ Reynolds shear stress profile. The points (colour online) denote the corresponding lift-included 1WC PP-DNS from \cite{costa2020corr}. Here $U_b$ is the flow bulk velocity.}
\label{fig:validation_pp}
\end{figure}
The present code has demonstrated successful applications in wall-bounded particle-laden flows, including open channel flow \citep{wang2019two,gao2023direct} and planar Couette flow \citep{richter2014,richter2015turbulence} loaded with inertial particles, but the lift force is neglected. Hence, for completeness, we present the validation of the one-way-coupling PP-DNS code, including the lift force models. Specifically, we reproduced in figure~\ref{fig:validation_pp} the PP-DNS of particle-laden turbulent channel flow in the one-way coupling regime using these three lift models, as reported in \cite{costa2020interface,costa2020corr} in the one-way coupling regime. This corresponds to case \texttt{CL1} in Table~\ref{tab:sum}. The agreement is excellent, which validates our implementation.

\subsection{Computational parameters}
\label{sec:np}
We perform point-particle DNS of channel flow at $\Rey_{\tau}=180$ in a computational domain $(L_x,L_y,L_z) = (6h,2h,3h)$, where $h$ is the half-channel height. 
The domain is discretized on $(N_x,N_y,N_z) = (160,320,160)$ grid points; the grid is slightly stretched to refine the near-wall resolution, corresponding to a grid spacing of $(\Delta_x^+,\Delta_y^+,\Delta_z^+) = (6.75,0.9,3.375)$, with the conventional `$+$' superscript denoting viscous wall scaling. Note that the grid spacing near the wall is comparable to that of the largest particle sizes. All simulations are carried out at a constant time-marching step fixed at $\Delta t^+ = 0.1$ (normalized by viscous unit $\nu/u^2_{\tau}$), which corresponds to $\text{CFL} \approx 0.4$. The DNS results confirm this time step to be sufficiently small that particles and fluid elements could not pass through a grid cell per time step \citep{zheng2021modulation}. The total simulation time is $T \approx 450h/u_\tau$ (about $80\,000$ viscous time scales $\nu/u_\tau^2$), which is long enough to guarantee converged statistics \citep{sardina2012wall}.

\begin{table}
  \begin{center}
\def~{\hphantom{0}}
\begin{tabular}{lccc}
       Case           & $D^+$  & $\rho_p/\rho$ & $N_p$\\[3pt]
       \texttt{CPF}   & \multicolumn{3}{c}{particle-free case}\\
       \texttt{CL1}   &  3   & 100 &   $5 \times 10^4$ \\
       \texttt{CM1}   &  1   & 900 &   $1.5 \times 10^5$ \\
       \texttt{CS1}   &  0.1 & 90000 &  $1.5 \times 10^6$
  \end{tabular}
  \caption{Physical parameters used in the present point-particle DNS campaign. $N_p$ denotes the total number of particles, and $D^+$ is inner-scaled particle diameter. The particle density ratio was adjusted to keep the same value of the Stokes number $St^+=\tau_{p}/(\nu/u_\tau^2) = 50$ and bulk solid mass fraction $\Psi_m = 0.337$ for all cases.}
  \label{tab:sum}
  \end{center}
\end{table}

We consider three different setups for varying particle diameter, while $St^+ = \tau_{p}/(\nu /u_\tau^2)=50$ is fixed by varying the particle density. These parameters are described in table~\ref{tab:sum}. This target particle Stokes number was chosen since it is known to feature highly inhomogeneous particle distributions in wall turbulence; see \cite{sardina2012wall}. Moreover, it should be noted that a value of $D^+ =3$ corresponds to about one Kolmogorov length scale in the channel bulk, an order of magnitude which is often investigated in the literature using PP-DNS \cite[see, e.g.,][]{bernardini2014reynolds,motoori2022role,zhang2022dominant}. We performed one- and two-way coupling simulations at a fixed mass fraction for different particle sizes and lift force models. Naturally, in the one-way coupled simulations, the mass fraction is not a governing parameter, and the high number of particles ensures statistical convergence of the results. In the two-way coupling PP-DNS, the bulk mass fraction is fixed to about $\Psi_m \approx 30\%$, to ensure significant turbulence modulation and highlight the effects of lift force model choice in the flow statistics.

\section{Results and discussion}
\label{sec:res}

\subsection{Particle dynamics}
\label{sec:1wc}

We start by describing the dynamics of the dispersed phase under the effect of lift force models using a simple, one-way coupling approach. This allows us to measure observables sampled by the particle that dictate the validity region of the lift models (e.g., $Sr$, $\varepsilon$, and $\Rey_p$). As we will see, current lift models are bound to fail in the buffer layer.

\begin{figure}
\centerline{\includegraphics[width=1.0\textwidth,keepaspectratio]{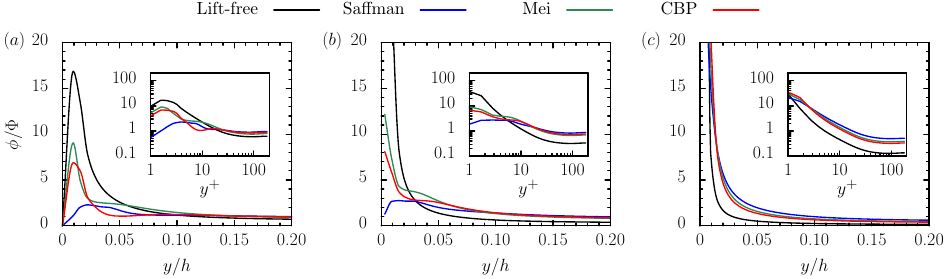}}
\caption{Local particle mass fraction $\phi$, normalized by the bulk value $\Phi$ as a function of the wall-normal distance (1WC). $(a)$ \texttt{CL1}, $(b)$ \texttt{CM1}, $(c)$ \texttt{CS1}. The inset shows the normalized particle volume fraction, versus the inner-scaled wall distance.}
\label{fig:phiv_1wc}
\end{figure}

Figure~\ref{fig:phiv_1wc} shows the profiles of the normalized mean particle concentration for all particle sizes. When the lift force is not taken into account, the preferential accumulation of particles in the wall region is significant, which has been widely observed \citep{picano2009spatial,sardina2012wall,lee2015modification,gao2023direct,Gualtieri2023effect} and termed as turbophoresis \citep{reeks1983transport,johnson2020turbophoresis}. However, the particles show less wall accumulation when subjected to lift forces, and this effect is more pronounced the larger the particle size. The Saffman model yields a much smaller peak, while the other two models feature smaller deviations, with the highest peak corresponding to the Mei model. When the particles are very small ($D^+=0.1$), the differences in the lift force models become almost indistinguishable, with all the models robustly showing the same reduction of wall accumulation. Yet the difference in accumulation in the absence of lift is clear.

\begin{figure}
\centerline{\includegraphics[width=0.8\textwidth,keepaspectratio]{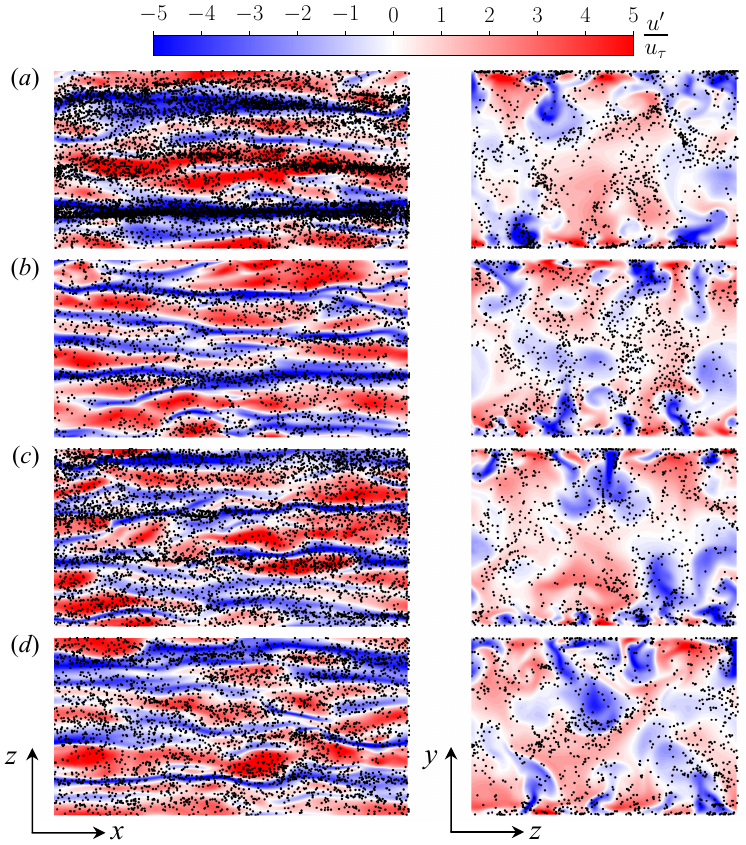}}
\caption{Inner-scaled instantaneous streamwise velocity fluctuation and the corresponding instantaneous snapshots of particle locations in $x$-$z$ ($y^+=12$, left column) and $y$-$z$ ($x^+=540$, right column) planes (Case \texttt{CL1}, 1WC). The wall-parallel plane shows all particles with inner-scaled wall normal position $Y_p^+ \le 12$. $(a)$ Lift-free, $(b)$ Saffman model, $(c)$ Mei model, $(d)$ CBP model.}
\label{fig:visu}
\end{figure}

\begin{figure}
\centerline{\includegraphics[width=1.0\textwidth,keepaspectratio]{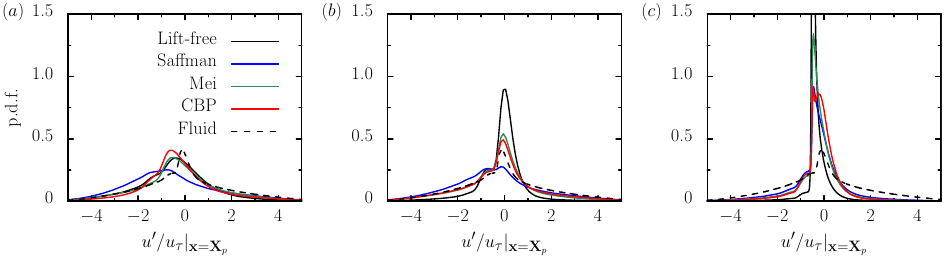}}
\caption{Probability density functions (p.d.f.s) of inner-scaled streamwise fluid velocity fluctuations conditioned at the particle location within the $y^+ \le 12$ region (1WC).
$(a)$ \texttt{CL1}, $(b)$ \texttt{CM1}, $(c)$ \texttt{CS1}.}
\label{fig:pdf_uf}
\end{figure}

To qualitatively describe the lift force effect on the spatial localization of wall accumulation, figure~\ref{fig:visu} shows instantaneous snapshots of streamwise velocity fluctuations in a wall-parallel plane within the buffer layer ($y^+=12$) and in a plane of constant streamwise location ($x^+=540$), along with the corresponding particle positions for the case with largest particles \texttt{CL1}.
As expected, the particles show strong spatial localization and inhomogeneous distribution, with larger local density occurring in elongated clusters \citep{sardina2012wall}. Effects of the lift force model on the gross characteristics of the particle distribution are readily discernible. One notable manifestation is the difference in particle numbers in the wall-parallel ($x$-$z$) plane. Consistent with figure~\ref{fig:phiv_1wc}, the lift-free model yields the highest number of particles, while the Saffman model yields the lowest. Conversely, the Mei and CBP models fall somewhere between these two. This indicates the weakening accumulation of particles near the wall by the lift force, consistent with observations by \cite{marchioli2007influence} and \cite{shin2022dynamics}. The other manifestation concerns small-scale particle clustering. At first glance, lift-free, Mei and CBP simulations show a more significant tendency to over-sample low-speed regions than the Saffman one. This tendency is checked in figure~\ref{fig:pdf_uf}, showing the probability density functions (p.d.f.s) of streamwise fluid velocity fluctuations sampled by the particles within the $y^+ \le 12$ region. For larger particle sizes, particles under the Saffman lift force sample a less focused range of streamwise velocity fluctuations than other models for larger particle sizes. In particular, for $D^+=3$ and $1$, the CBP and Mei cases show about the same tendency to sample low-speed regions, differently than the Saffman model and comparable to the lift-free case for $D^+=3$. Preferential accumulation gets more focused with decreasing particle size, with the lift-free case producing the strongest preferential accumulation for $D^+=1$ and $0.1$.

\begin{figure}
\centerline{\includegraphics[width=1.0\textwidth,keepaspectratio]{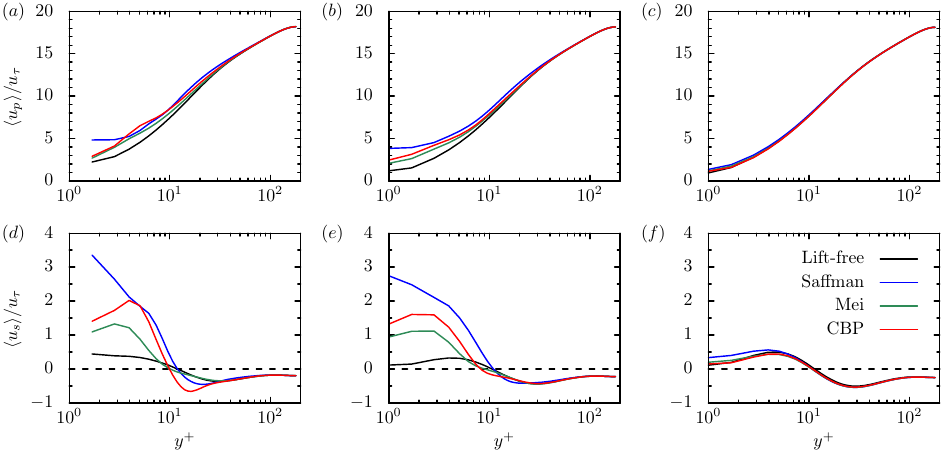}}
\caption{Inner-scaled streamwise mean particle velocity and slip velocity profiles (1WC). $(a,d)$ \texttt{CL1}, $(b,e)$ \texttt{CM1}, $(c,f)$ \texttt{CS1}. The horizontal dashed lines denote $\langle u_s \rangle = 0$.}
\label{fig:up_1wc}
\end{figure}

\begin{figure}
\centerline{\includegraphics[width=1.0\textwidth,keepaspectratio]{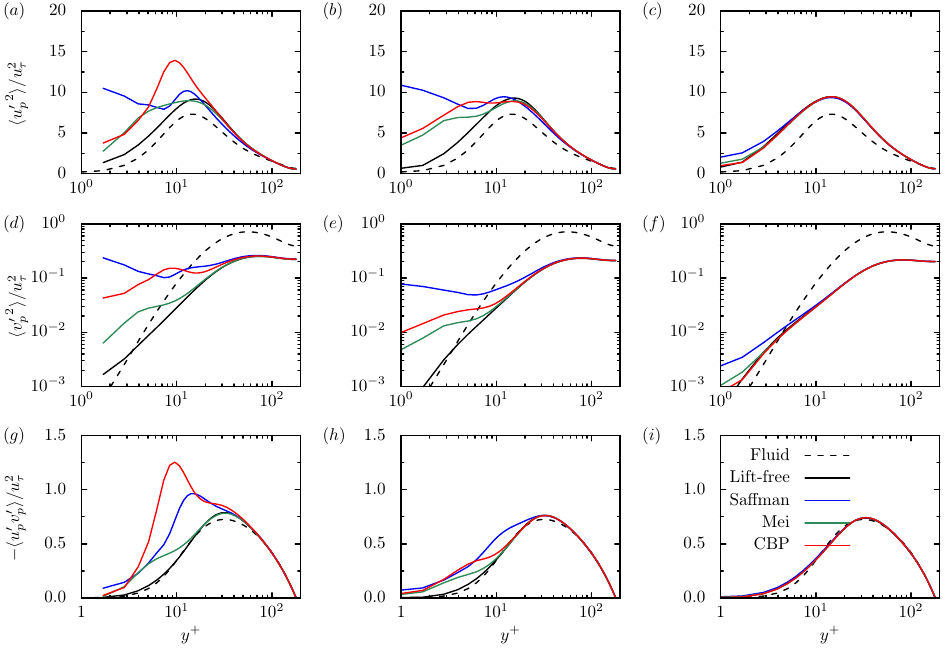}}
\caption{Second-order moments of mean particle (solid lines) and fluid (dashed lines) velocity (1WC). $(a,b,c)$ Inner-scaled streamwise and $(d,e,f)$ wall-normal velocity variances, and $(g,h,i)$ Reynolds shear stress profiles. The left $(a,d,g)$, middle $(b,e,h)$ and right $(c,f,i)$ columns denote Case \texttt{CL1}, \texttt{CM1} and \texttt{CS1}, respectively.}
\label{fig:prms_1wc}
\end{figure}

Figure~\ref{fig:up_1wc} shows the inner-scaled profiles of streamwise mean particle velocity $\langle u_p \rangle$ and local slip velocity $\langle u_s \rangle$. Compared with the lift-free particles with $D^+ \ge 1$, lift forces increase $\langle u_p \rangle$ in the viscous sublayer and buffer layer, while the differences between cases become negligible beyond $y^+ \sim 30$. Once more, particles under the Saffman lift force show the highest variation with respect to the lift-free case, yielding the highest $\langle u_p \rangle$; particles under the Mei lift force yield the lowest, and the ones under the CBP lift force fall in between. Consistently, for the smallest particle sizes, the overall modification of $\langle u_p \rangle$ by lift forces is quite small. This tendency of lift forces becoming less important with decreasing particle size is expected -- as we will discuss later in more detail (\S\ref{sec:challenges}), the relative importance of lift to drag forces close to the wall roughly scales with the inner-scaled particle diameter $D^+$.

At the wall, while the fluid velocity must vanish, particles can feature a mean apparent slip where they flow faster than the fluid due to their inertia \citep{zhao2012stokes}. This is clearly shown in the particle mean slip (figure~\ref{fig:up_1wc}$d,e,f$). At larger wall distances, the tendency of particles to over-sample slower-than-average fluid velocity \citep{kiger2002suspension,baker2021particle} is also clearly reflected in the mean particle slip velocity. A similar trend can be observed in the corresponding streamwise mean slip velocity profiles; however, the difference in $\langle u_s \rangle$ is larger than $\langle u_p \rangle$, and this is attributed to the negative apparent slip velocity ($\langle u |_{\mathbf{x}=\mathbf{X}_p} \rangle - \langle u \rangle < 0$), which reflects preferential sampling of slower-than-average fluid. Interestingly, lift forces increase the apparent slip velocity near the wall. This can be understood from the sign of the lift force in this region, which points towards the wall due to the positive slip velocity. Particles sampling higher-momentum regions will be driven towards the wall by the lift force, and their inertia results in a higher mean slip.

The sign of $\langle u_s \rangle$ changes in roughly the same location of the buffer layer, irrespective of the particle size and lift force model. This mechanism seems relatively robust and has been observed in numerous other studies with inertial particles at different Stokes numbers, and different flow Reynolds numbers \cite[see, e.g.,][]{mortensen2008dynamics,zhao2012stokes,wang2020multiscale,gao2023direct}. It will be shown that this change in sign makes lift force modelling in wall turbulence extremely challenging (see discussions in \S\ref{sec:challenges}).

Figure~\ref{fig:prms_1wc} shows the $x-y$ components inner-scaled velocity covariances for the particle and fluid phases, respectively. The spanwise velocity variances are hardly affected by lift forces, and thus not shown. As expected, away from the wall where the mean shear is low, the particle statistics closely follow those of the lift-free cases. Near the wall, instead, the lift force enhances the streamwise and wall-normal particle velocity variances ($\langle u'^2_p \rangle$ and $\langle v'^2_p \rangle$) and Reynolds stress ($\langle u'_p v'_p \rangle$), even for the smallest particle size ($D^+=0.1$). 
Enhancement of wall-normal velocity fluctuations near the wall has indeed been reported in previous numerical and experimental studies \citep{fong2019velocity,costa2020interface}. Indeed, as also shown in \cite{costa2020interface}, this enhancement of velocity fluctuations near the wall cannot be reproduced when lift forces are not considered in the particle dynamics. It seems that lift force is very effective at correlating streamwise and wall-normal velocity fluctuations near the wall -- under high shear rate, small variations in streamwise velocity naturally translate to changes in vertical acceleration through the lift forces, which in turn induce wall-normal velocity fluctuations. Streamwise velocity fluctuations will also be naturally amplified, since streamwise and wall-normal velocity fluctuations are correlated through the mean shear. As we will see in \S\ref{subsec:2wc}, this amplification of near-wall velocity fluctuations by lift forces has major consequences on turbulent modulation and drag changes in the flow at high particle mass loading. 

\subsection{Mechanism for near-wall accumulation}
\label{sec:nwa}
\begin{figure}
\centerline{\includegraphics[width=0.65\textwidth,keepaspectratio]{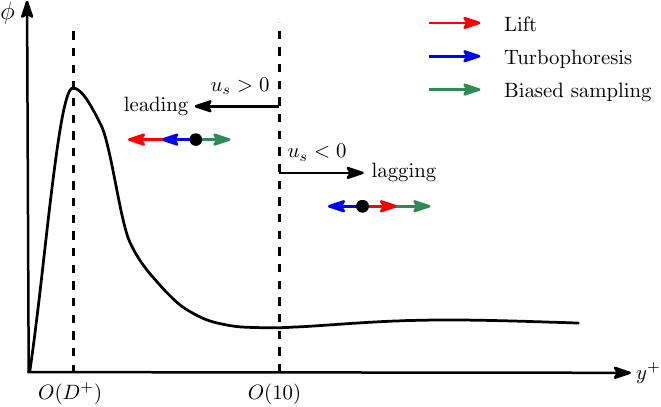}}
\caption{Illustration of the interplay between different mechanisms responsible for the wall-normal particle transport.}
\label{fig:mech}
\end{figure}

To understand the role of the lift force in altering the near-wall particle distributions, it is essential to examine three primary mechanisms responsible for wall-normal inertial particle transport. The first is the lift-induced migration, which is the direct consequence of the inertial lift force. The second is turbophoresis \citep{reeks1983transport}, causing particle migration from regions of higher to lower turbulence intensity. In wall-bounded flows, turbophoresis results in strong particle accumulation in the viscous sublayer, where turbulent fluctuations vanish. Note however that in most cases the peak in the particle concentration appears not at the wall but at a distance of $O(D)$ from the wall \citep{marchioli2002mechanisms,costa2020interface}, owing to the wall--particle collisions as well as hydrodynamic wall--particle interactions \citep{goldman1967slow,vasseur1977lateral,zeng2005wall}. Finally, there is a biased sampling effect, due to the tendency of particles to sample ejection regions in the buffer layer featuring negative streamwise velocity and positive (i.e. repelling) wall-normal velocity fluctuations \citep{marchioli2002mechanisms,sardina2012wall}. As a result, biased sampling usually provides a net slow particle drift away from the wall. The interaction among the three mechanisms is sketched in figure \ref{fig:mech} and will be elaborated in the following.

Lift-induced migration is directed along $\boldsymbol{\omega}\times\mathbf{U}_s$, which in the case of a turbulent wall flow is predominantly in the wall-normal direction. According to figure \ref{fig:up_1wc}($d-f$),  particles lead the liquid flow for $y^+\lesssim10$ and lag behind the flow further away from the wall. Hence the lift-induced migration is \emph{towards} the wall in the viscous sublayer, and acts cooperatively with the turbophoresis effect. The way these two mechanisms interact in the viscous sublayer may seem surprising at first glance, as figure \ref{fig:phiv_1wc} clearly indicates a suppression of the near-wall particle accumulation by lift force. The reasoning behind this suppression may be understood by noting the following two issues. First, the lift force is towards the channel centre within the buffer layer and beyond $y^+\approx10$. The resulting outward migration, together with the strong turbulence intensity in the buffer layer, tends to entrain particles in this region outwards, leading to a net decrease in the concentration in the region $y^+\lesssim 10$ as revealed from figure \ref{fig:phiv_1wc}. Second, the lift force influences the wall-normal particle transport by turbophoresis. This phenomenon is linked to the particle wall-normal velocity variance $\langle v'^2_p \rangle$ \citep{reeks1983transport,johnson2020turbophoresis}. Indeed, in the limit of vanishing particle Reynolds number, the particle wall-normal momentum balance at steady state yields a turbophoresis pseudo-force proportional to $\mathrm{d}\langle v'^2_p \rangle/\mathrm{d}y$, which creates the migration of particles down gradients in particle wall-normal velocity variance \citep{sikovsky2014singularity,johnson2020turbophoresis}. Recall figure~\ref{fig:prms_1wc}($d,e,f$), where the variation of $\langle v'^2_p \rangle$ with the wall distance is depicted. Clearly, the lift force tends to increase the ``apparent inertia'' of the particles, enhancing the deviation between the particle velocity variance and that of the fluid within the viscous sublayer. This leads to an attenuation in the corresponding turbophoresis pseudo-force. In particular, for the two cases where $D^+\geq1$ (figure \ref{fig:prms_1wc}$d,e$), the prediction with the Saffman lift model yields $\mathrm{d}\langle v'^2_p \rangle/\mathrm{d}y<0$ in the viscous sublayer, indicating that the turbophoresis pseudo-force is outwards, driving the particles away from the wall. Consequently, the lift force competes with the turbophoresis pseudo-force in these two cases, leading to a highly flattened near-wall peak in the particle fraction as seen in figure \ref{fig:phiv_1wc}($a,b$). In contrast, they cooperate in predictions using the Mei and the CBP models, as figure \ref{fig:phiv_1wc}($a$) shows. There, the lift force compensates the attenuation in turbophoresis, making the negative slopes in $\phi(y^+)$ from results using Mei and the CBP models closely follow the lift-free case (highlighted in the figure insets).
Note, however, that the magnitude of the Mei lift (CBP lift) is proportional to $Re_\omega^{1/2}$ ($Re_\omega$), which approaches $D^+$ (${D^+}^2$) in the viscous sublayer. Consequently, for the two cases where $D^+\leq1$ the lift-induced migration is weak and incapable of compensating the still-pronounced attenuation in turbophoresis revealed from figure \ref{fig:prms_1wc}$(e,f)$. This is why in these two cases the slopes in $\phi(y^+)$ from results employing the Mei and CBP models are smaller in magnitude than that in the lift-free case. Still, the difference in accumulation with respect to the lift-free case for smaller sizes is clear, due to near-wall particles residing for long times near the wall while being subjected to small but persistent lift forces.

\begin{figure}
\centerline{\includegraphics[width=1.0\textwidth,keepaspectratio]{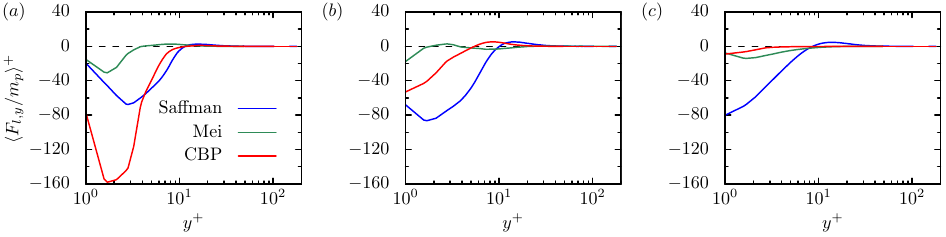}}
\caption{The inner-scaled lift-induced acceleration (i.e., normalized by $u^3_{\utau}/\nu$) profiles along the wall-normal direction (1WC). $(a)$ \texttt{CL1}, $(b)$ \texttt{CM1}, $(c)$ \texttt{CS1}. The horizontal dashed lines denote $\langle F_{l,y} \rangle = 0$.}
\label{fig:lift_1wc}
\end{figure}

Given the discussion above, it is worth examining in more detail the link between the chosen lift model and the corresponding modulation in the turbophoresis pseudo-force, particularly whether to what extent this modulation relates to the lift force magnitude. Combing the turbophoresis pseudo-force (figure~\ref{fig:prms_1wc}$d,e,f$) and lift force as shown in figure \ref{fig:lift_1wc}, a monotonic increase in the suppression of the turbophoresis pseudo-force with increasing lift is observed at $D^+ \leq 1$, while no such trend is evident at $D^+=3$. In the latter case, the suppression of the turbophoresis pseudo-force is most pronounced in the prediction using the Saffman model, even though the magnitude of the corresponding lift force is only half that of the prediction using the CBP model for $y^+\lesssim5$.
Hence, the differences in peak concentration are directly linked not to the lift force magnitudes, but rather to the modulation in the turbophoresis pseudo-force induced by these models. In the present work, the turbophoresis pseudo-force is always suppressed by the presence of lift, leading to a decrease in the peak concentration. Hence, the precise relationship between the extent of this suppression and the lift force magnitude remains unclear, making it challenging to directly connect the chosen lift models with the changes in the predicted concentration profiles. Finally, it would be worth investigating these dynamics at lower Stokes numbers, as both turbophoresis and preferential sweeping will be less pronounced, and the nature of the interaction between lift and turbulence may drastically change. Given the increased issues with lift force modelling at lower Stokes number, discussed in the next section, addressing this question should be the object of a future, dedicated study.

To gain insights into biased sampling of sweeps/ejection events under lift forces, we carried out a quadrant analysis for the fluid flow experienced by the particles. In the ($u^{\prime}-v^{\prime}$) plane, with positive $v^{\prime}$ directed away from the wall, the flow experienced by the particle is categorized into four types of events: first quadrant events (Q1), characterized by outward motion of high-speed fluid, with $u^{\prime} > 0$ and $v^{\prime} > 0$; second quadrant events (Q2), characterized by outward motion of low-speed fluid, with $u^{\prime} < 0$ and $v^{\prime} > 0$, which are usually called ejections; third quadrant events (Q3), characterized by inward motion of low-speed fluid, with $u^{\prime} < 0$ and $v^{\prime} < 0$; and finally, fourth quadrant events (Q4), which represent motions of high-speed fluid towards the wall, with $u^{\prime} > 0$ and $v^{\prime} < 0$, and are usually called sweeps. 
\begin{figure}
\centerline{\includegraphics[width=1.0\textwidth,keepaspectratio]{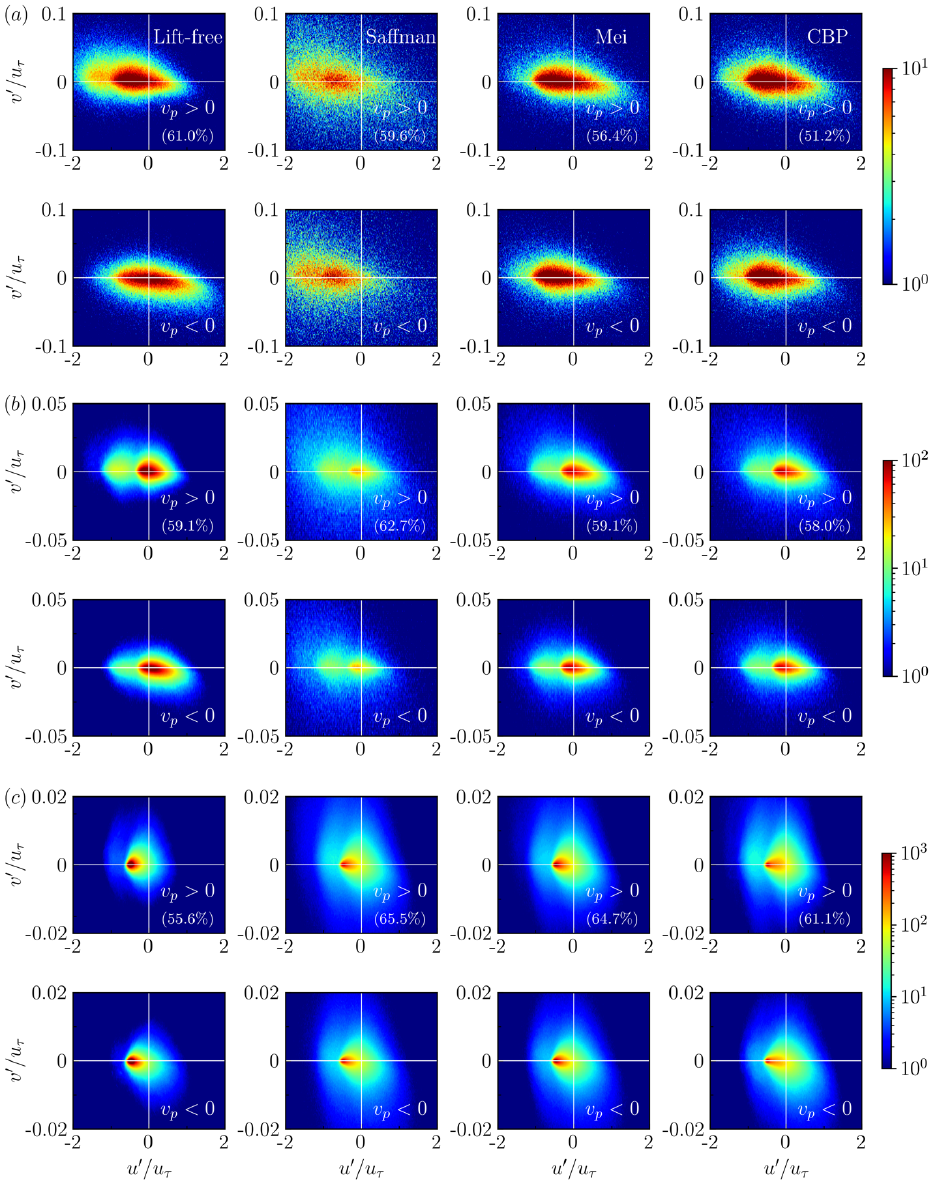}}
\caption{Joint p.d.f.s of the streamwise and wall-normal fluid fluctuating velocities seen by ascending ($v_p >0$) and descending ($v_p <0$) particles within the $y^+ \le 12$ region (1WC).
 $(a)$ \texttt{CL1}, $(b)$ \texttt{CM1}, $(c)$ \texttt{CS1}. The numbers in the panels denote the proportion of ascending particles.}
\label{fig:quadrant_pdf_dp}
\end{figure}
Figure \ref{fig:quadrant_pdf_dp} shows the joint p.d.f.s in the ($u^{\prime}-v^{\prime}$) plane of the fluid fluctuation seen by ascending ($v_p >0$) and descending ($v_p <0$) particles within the $y^+ \le 12$ region. The first observation is that a larger proportion of particles tends to move outwards (ascending), irrespective of the presence of lift force. The proportion of ascending particles without lift force decreases as the particle size decreases, whereas particles subject to lift force exhibit an inverse trend. Second, the lift-free ascending particles are highly likely to sample ejections (Q2), while the lift-free descending ones are prone to sample sweeps (Q4) for particles with $D^+ \geq 1$. This agrees with experimental observations in channel flow laden with finite-sized particles \citep{kiger2002suspension,yu2016finite,baker2021particle}.
However, the particles under the effect of lift forces tend not to over-sample sweeps, irrespective of the wall-normal particle velocity sign, which indicates that the wall-normal velocity of the particles is less correlated with the fluid one. This also confirms that inertial particles under the effect of lift forces are not prone to slowly drive towards the wall in low-speed regions, because the competing wall-repelling effect of lift forces seems to dominate their dynamics. In close inspection of the high-intensity region of the joint p.d.f.s on the $u'$-axis, we note that particles under the Saffman lift force tend to sample a wider range of negative $u'$ regions, compared with particles under the Mei and CBP lift forces. This aligns with the p.d.f.s of $u'$ as shown in figure~\ref{fig:pdf_uf}. Third, irrespective of the presence of the lift force or wall-normal particle velocity sign, the smallest particles ($D^+=0.1$) over-sample ejection (Q2) and inward motions of low-speed fluid regions (Q3). Sweep events, bringing high-momentum fluid towards the wall are only weakly experienced by particles. 
Overall, the percentage of particles sampling Q2 events only slightly surpasses that for Q3, suggesting that biased sampling may play a minor role in driving high inertial particles ($St^+=50$) to depart from the inner wall region. This observation aligns with numerical results from \citet{marchioli2002mechanisms}, where biased sampling is found to be more efficient for transferring particles with smaller inertia.

\subsection{Emerging challenges in lift force modelling}
\label{sec:challenges}

The aforementioned results allow us to address several issues concerning the effects of lift force in turbulent wall flows, which may have been previously overlooked in PP-DNS simulations. Ideally, such discussions would be based on results from PR-DNS simulations. However, in this instance, we will extrapolate from PP-DNS results, in conjunction with the CBP model, as an analogue to PR-DNS outcomes. This approach is justified by findings from \citet{costa2020interface,costa2020corr}, which demonstrate that PP-DNS simulations are capable of satisfactorily reproducing the corresponding PR-DNS results for the most challenging case considered in this study, with $D^+=3$.

Let us first address the circumstances under which the effects of lift force could be considered negligible for large inertial particles, and thereby disregarded in PP-DNS simulations. Given that biased sampling does not play a key role in the wall-normal transport of inertial particles, a criterion based on the ratio of lift to the turbophoresis pseudo-force seems sufficient to describe the relative significance of the lift force. In this context, the dimensionless particle size $D^+$ seems a promising candidate, owing to the following two reasons. First, ${D^+}^2$ approaches the shear Reynolds number $Re_\omega$ in the viscous sublayer; the latter directly measures the magnitude of the shear lift irrespective of the choice of lift models \citep{saffman1965lift,mclaughlin1991inertial,mei1992approximate,shi2019lift,costa2020corr}. Second, the turbophoresis effects seem to decay with increasing $D^+$. This is clearly revealed by comparing the slopes of $\langle v'^2_p \rangle(y^+)$ revealed from figure \ref{fig:prms_1wc}$(d,e,f)$ which correspond to results with increasing $D^+$. Glueing these two aspects together, it seems that the force ratio of lift to turbophoresis roughly scales as ${D^+}^k$ with $k>0$. Although the exact value of $k$ is infeasible to be determined, the results summarized in \S\ref{sec:1wc} made it clear that by no means should the lift force be neglected for $D^+\geq1$. On the other hand, it is reasonable for the lift force to be neglected for $D^+\leq0.1$, except for the underprediction of near-wall accumulation discussed above (recall figure~\ref{fig:phiv_1wc}).

\begin{figure}
\centerline{\includegraphics[width=1.0\textwidth,keepaspectratio]{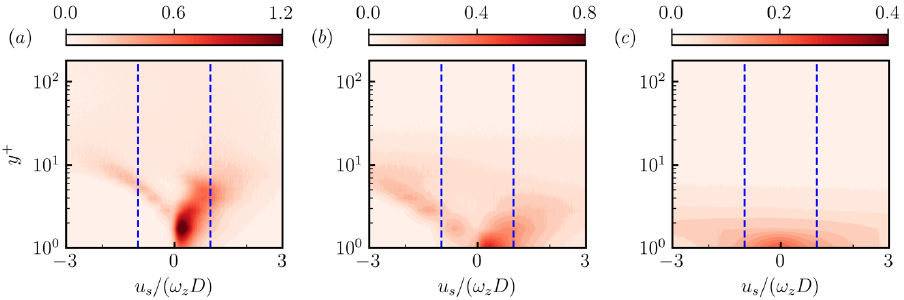}}
\caption{The p.d.f.s of  $u_s/(\omega_z D)$ along the wall-normal direction (1WC DNS with the CBP model). $(a)$ \texttt{CL1}, $(b)$ \texttt{CM1}, $(c)$ \texttt{CS1}. The vertical dashed lines denote $u_s/(\omega_z D) = \pm 1$.}
\label{fig:pdf_usomgD_dp}
\end{figure}

We should now address a major issue in commonly applied lift force models for PP-DNS simulations of turbulent wall flows. As outlined in section \ref{sec:model}, any of the shear-lift models originating from the pioneering work of \citet{saffman1965lift} are rigorously applicable only in the double limits $\Rey_p\to 0$ and $\Rey_\omega\to 0$. In the viscous sublayer, $\Rey_p\approx D^+(\langle u_s\rangle/u_\tau)\approx D^+$ for $D^+\geq1$ according to figure \ref{fig:up_1wc}, whereas $\Rey_\omega\approx {D^+}^2$. Consequently, neither of the two limits is satisfied for $D^+\geq1$. In addition to the double limits above, there is a constraint concerning the dimensionless shear rate $Sr$. This is made clear by noting that $Sr=\Rey_\omega^{1/2}\varepsilon$ and hence $Sr\leq O(1)$ even in the strong shear limit, $\varepsilon\geq1$, considered in \citet{saffman1965lift}. This is why in the wall-bounded situation the ``outer-region" lift solutions \citep{asmolov1990dynamics,mclaughlin1993lift,takemura2009migration} following the methodology of \citet{saffman1965lift} match with the corresponding ``inner-region" solutions \citep{cherukat1994inertial,magnaudet2003drag,shi2020lift} only for $Sr=O(1)$. For particle-laden channel flows with small inertial particles \citep{mortensen2008dynamics,zhao2012stokes,wang2020multiscale,gao2023direct}, this constraint is more seriously violated than that for the double limits above, as the slip velocity reverses at $y^+\approx O(10)$ (see, e.g., figure \ref{fig:up_1wc}$(d,e,f)$ in the present work), making $Sr\to\infty$ for particles within the buffer layer. This issue is highlighted in figure \ref{fig:pdf_usomgD_dp}, where we show the p.d.f.s of the inverse of $Sr$ (horizontal axis) as a function of the wall-normal distance $y^+$ (vertical axis). Apparently, a large portion of particles experiences a strong shear where $Sr\geq1$, in particular for the two cases where $D^+\geq1$. Finally, it is worth noting that the probability of a particle experiencing very high values of $Sr$ near the wall will increase with decreasing values of $St^+$ as, for the same local shear rate, particles will experience an ever smaller slip velocity. This exacerbates the violation of the $Sr=O(1)$ constraint, meaning that current lift force models are even less suitable for turbulent channel flow laden with particles at lower Stokes numbers.

Additionally, the discussion above focuses on the lift force arising from the shear in the ambient flow. In wall-bounded flows, as considered in this work, the lift force may deviate significantly from its unbounded counterpart if the particle is close to the wall. In brief, the presence of the wall leads to a repulsive transverse force in the absence of the ambient shear \citep{vasseur1977lateral,zeng2005wall} and, in the presence of an ambient shear, a suppression of the shear-induced lift \citep{mclaughlin1993lift,magnaudet2003drag,takemura2009migration}. Although analytical solutions for the combined lift force are achievable in some asymptotic limits (see, e.g., \citet{wang1997role} and, more recently, \cite{shi2020lift} for an overview of these solutions), there is so far no satisfactory way to glue these solutions together to achieve a general treatment of the wall effect \citep{ekanayake2020lift,ekanayake2021lift}. Here, instead of trying to make any further progress in achieving such a general solution, it is more feasible to discuss under which condition may this intricate wall effect be neglected. Within the $O(D)$-depth layer close to the wall, wall--particle collisions govern the wall-normal position of the peak particle concentration. Hence, the hydrodynamic wall--particle interaction may play a significant role only beyond $y^+\approx D^+$. On the other hand, it is known that \citep{mclaughlin1993lift} the amount of reduction of the shear-induced lift due to the wall decreases as $(\ell/\ell_\omega)^{-5/3}$ and reduces to only $20\%$ of its magnitude in the limit $\ell/\ell_\omega\to0$ at $\ell/\ell_\omega\approx3$. Hence, the wall effect may be considered negligible if the position of the peak particle concentration, which is of $O(D)$ from the wall, is larger than about $3\ell_\omega$, which yields roughly $D^+\geq3$. This finding, together with the previous one where (shear-induced) lift force is non-negligible for $D^+\geq1$, indicates that such a wall effect can be disregarded except in the overlap regime where $1\leq D^+ \leq3$.

In addition to the ambient shear, particle rotation is known to produce a lift force as well \citep{rubinow1961transverse}. This lift contribution is often neglected in prior PP-DNS simulations and not considered in the present work. Nevertheless, it might be worth discussing in the following its relevant importance. Assuming that the particle inertia is small enough for the particle to always stay in the torque-free state, particle rotation leads to a lift contribution of $C_L^\Omega=0.5Sr$ in the double limits considered by \citet{saffman1965lift}. The ratio between this spin-induced lift contribution and that induced by ambient shear is proportional to $Re_\omega^{1/2}$, indicating that $C_L^\Omega$ is merely a component of the second-order lift contribution from the ambient shear. This is confirmed by recent work of \citet{candelier_mehaddi_mehlig_magnaudet_2023}, where second-order lift contributions (with respect to $Re_\omega^{1/2}$) were obtained using matched asymptotic expansions. In addition to the spin-induced contribution $C_L^\Omega=0.5Sr$, they also obtained a second-order contribution from the ambient shear $C_L^{'(2)}=-0.505Sr$, which counterbalances the spin-induced contribution. In other words, the effect of particle rotation on the lift is negligible in the torque-free state. This conclusion is valid irrespective of the presence of the wall \citep{cherukat1994inertial,magnaudet2003drag,shi2020lift}. However, large-inertia particles with a large density ratio as considered in this work might not always stay torque-free. Specifically, for particles with density ratio $\rho_p/\rho\geq100$, the relaxation time in response to torque is comparable with that to slip (see, e.g., \citet{bagchi2002effect}). This finding, together with the pronounced slip in the viscous sublayer revealed from figure \ref{fig:up_1wc}$(a,b)$, indicates that particles of $D^+\geq 1$ may experience a strong angular acceleration and possibly a non-negligible lift contribution from forced rotation. Investigating the effects of rotation by accounting for the particles' conservation of angular momentum would therefore be interesting but is left outside the scope of the present work due to the higher complexity associated with drag and lift closures for angular momentum.

Finally, it is worth noting that while the CBP lift force model used in \cite{costa2020interface,costa2020corr} appears to provide reasonable predictions in some cases, it is not universally valid. Specifically, it does not align with any of the inner-region solutions \citep{shi2020lift}, which are applicable for particles extremely close to the wall (e.g., within the viscous sublayer), nor does it approach the unbounded solution \citep{saffman1965lift, mclaughlin1991inertial} for particles located within the outer layer of the channel flow. The agreement between the PP-DNS simulations utilizing this lift force model and PR-DNS suggests that the model offers a satisfactory blending of results in the two limits for the specific cases explored in \cite{costa2020interface} (see table~\ref{tab:sum}). A preliminary attempt to establish a more general lift force model can be found in the work of \cite{ekanayake2020lift,ekanayake2021lift}, where a lift model satisfying the above conditions was proposed in the limit of small shear Reynolds numbers, i.e. $\Rey_\omega \ll1$. However, the applicability of this model is limited to $D^+\ll 1$ as, in the viscous layer, $\Rey_\omega \approx (u_\tau^2/\nu) D^2/\nu=(D^+)^2$. While this provides some insights, its practicality may be limited since the lift force no longer plays a pivotal role for $D^+<1$, as demonstrated here. Hence, further work extending the work of \cite{ekanayake2020lift,ekanayake2021lift} to finite $\Rey_\omega$ (hence $D^+$) would be significant.

\subsection{Lift-induced turbulence modulation}
\label{subsec:2wc}

Finally, in this section, we use two-way coupling point-particle DNS to illustrate that lift force models can have tremendous consequences in basic integral quantities such as the overall drag. To achieve this, we use the same parameters shown in table \ref{tab:sum}, which fixes the total mass fraction and particle Stokes number while varying the particle diameter. We should reiterate that high-fidelity two-way coupling point-particle DNS is highly challenging and is currently being actively investigated. Still, the simple particle-in-cell 2WC DNS employed here suffice to illustrate the impact of lift force model choice on important turbulence metrics. In what follows, we illustrate the effects of lift force on the overall drag and, analyze the streamwise momentum fluxes that contribute to it.

\begin{figure}
\centerline{\includegraphics[width=0.7\textwidth,keepaspectratio]{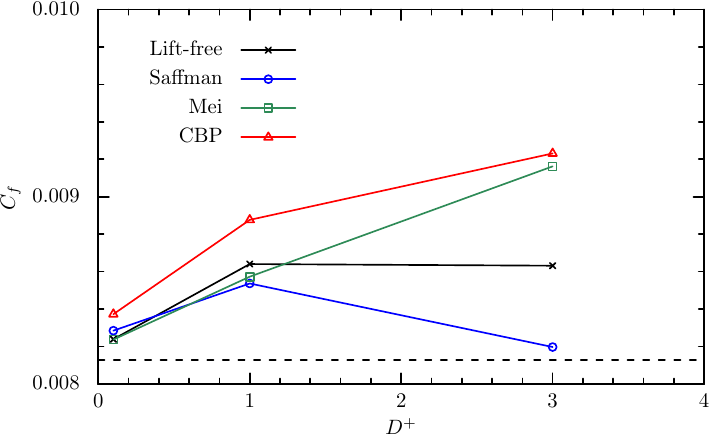}}
\caption{The wall friction coefficient $C_f$ versus the particle diameter $D^+$ (2WC). The horizontal dashed line denotes the mean $C_f$ of the particle-free case.}
\label{fig:cf_dp}
\end{figure}

Figure~\ref{fig:cf_dp} shows the friction coefficient for the 2WC DNS, defined as $C_f=2 u^2_{\tau}/U^2_b$. While all small inertial particles have a drag-increasing effect, the difference between lift force models is highly amplified as the particle size increases. At $D^+=3$, in particular, particles under the Saffman lift force show a drag reduction compared to the other particle sizes, almost reaching an overall drag decrease. Instead, the other lift models show a monotonic drag increase.

\begin{figure}
\centerline{\includegraphics[width=1.0\textwidth,keepaspectratio]{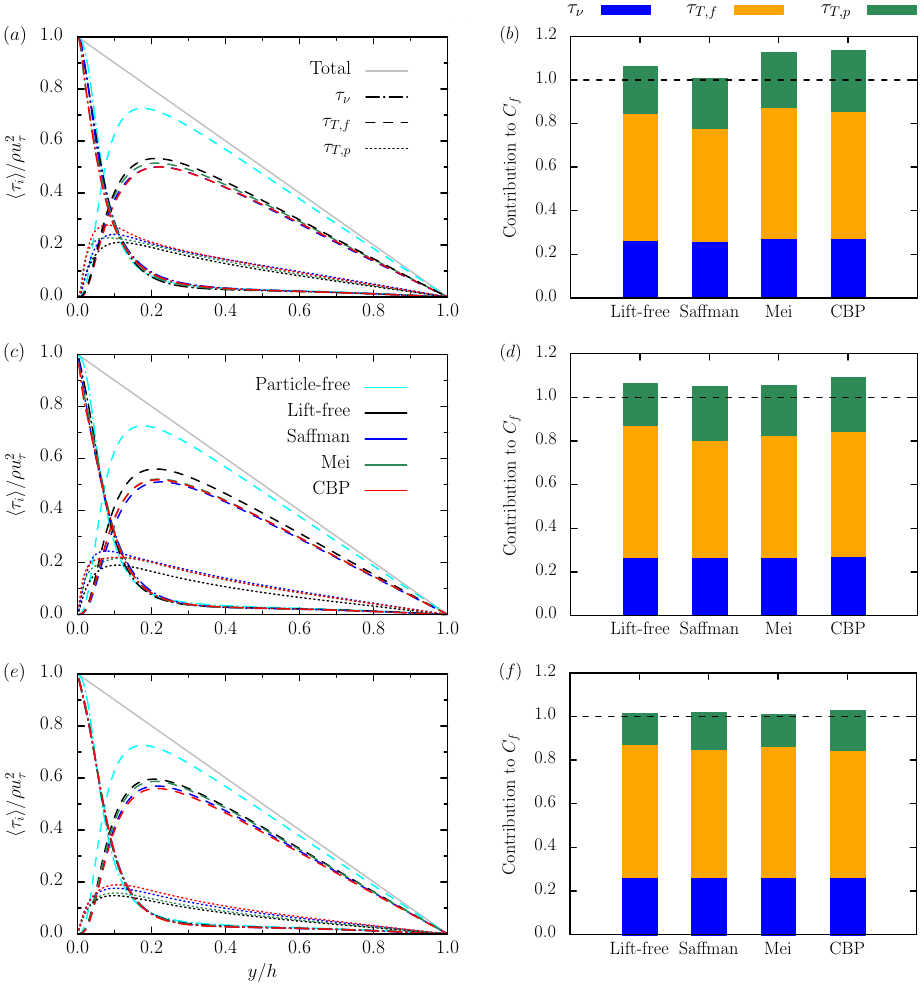}}
\caption{Budget of streamwise momentum $(a,c,e)$ and contribution of each stress contribution $\tau_i$ to the mean wall friction coefficient $C_f=2\tau_w/(\rho U_b^2)$ $(b,d,f)$, $1/(h^2)\int_0^h 6 (h-y) \tau_i \mathrm{d}y/(\rho U_b^2)$ normalized by that of the particle-free case.
$(a,b)$ \texttt{CL1}, $(c,d)$ \texttt{CM1}, $(e,f)$ \texttt{CS1}. Different line colours in $(a,c,e)$ denote different cases, which match the legend in $(c)$.}
\label{fig:budget}
\end{figure}

To better understand and quantify this difference, we follow the analysis in \cite{costa2021near} and investigate the stress budget in the two-way coupling limit of vanishing volume fraction, while keeping the mass fraction finite:
\begin{equation}
\tau = \rho u_\tau^2\left(1-\frac{y}{h}\right) \approx \underbrace{\left(\mu \frac{\mathrm{d}\left\langle u\right\rangle }{\mathrm{d}y} - \rho\left\langle u^\prime v^\prime\right\rangle\right)}_{\tau_\nu + \tau_{T,f}} - \underbrace{\left\langle \psi\right\rangle \left\langle u_p^\prime v_p^\prime\right\rangle}_{\normalfont\tau_{T,p}},
\label{eqn:two-way}
\end{equation}
where $\left<\psi\right> = \rho_p\left<\phi\right>$ is the mean local mass fraction. The terms in the right-hand side of this equation are the viscous stress, fluid Reynolds stress, and particle Reynolds stress (hereafter $\tau_\nu$, $\tau_{T,f}$, and $\tau_{T,p}$). This equation boils down to the stress budget in the single-phase limit ($\tau_\nu$ and $\tau_{T,f}$), corrected by an additional turbulent particle momentum flux, $\tau_{T,p}$ that involves the local mass fraction and the particle counterpart of the Reynolds shear stress. This term relates to the transfer of streamwise momentum by correlated particle velocity fluctuations, and shows the importance of high particle mass fraction near the wall when combined with correlated streamwise--wall-normal particle velocity fluctuations. We have seen in \S\ref{sec:1wc} that, while the former decreases in the presence of lift force, the latter may be increased by it.

The profiles of stress terms and the corresponding contributions of each momentum transfer mechanism to the overall friction coefficient, normalized by the overall friction coefficient of the particle-free case, are shown in figure~\ref{fig:budget}. On the left panels, the normalized shear stress profiles are shown, while on the right we show the so-called FIK identity employed to a particle suspension \citep{fukagata2002contribution,yu2021modulation}. In the case of a turbulent plane channel flow, this identity decomposes the mean friction coefficient ($C_f$) into a laminar and turbulent contribution. Here, the latter can be further decomposed into a contribution due to the fluid phase and one due to the particles. Each contribution is given by the following weighted integral $C_{f,i} = 1/h^2\int_0^h 6 (h-y) \tau_{i}\,\mathrm{d}y/(\rho U_b^2)$, with $\tau_i$ being one of the stress mechanisms described above. Note that the residual of this stress budget is virtually zero, which validates the approximation in equation~\eqref{eqn:two-way}.

As anticipated from figure~\ref{fig:cf_dp}, all 2WC DNS show increased drag compared to the unladen flow. Yet, remarkably, different lift models show qualitatively different turbulence modulation for the largest particle size ($D^+=3$). Particles under the Saffman lift force have a major net turbulence attenuation effect, which results in an almost drag-reducing flow with respect to the unladen case. Conversely, particles under the Mei and CBP lift forces have a net drag increase, mostly caused by an increase in the two-way coupling stress term $\psi\left<u'_pv_p'\right>$. This results from the amplification of particle Reynolds shear stresses $\left<u'_pv'_p\right>$ discussed in section \ref{sec:1wc}, which compensate for the decrease in local mass fraction $\psi$. Naturally, as the particle size decreases, the differences between lift models become less pronounced but still noticeable at all particle sizes, with particles under the Saffman lift force consistently showing the highest fluid turbulence attenuation. For the smallest particle size $D^+=0.1$, the two turbulent stresses (fluid and particle) terms still show noticeable but in practice negligible differences (recall  figure~\ref{fig:cf_dp}).

Hence, it appears that the most important dynamical effects of lift forces in two-way coupling conditions can be seen as finite-size effects that directly modify single-point moments of particle velocity for $D^+\gtrsim 1$ (recall figure~\ref{fig:prms_1wc}). However, the lift force still affects near-wall turbulence modulation when particles are relatively small, as they strongly affect the mean particle concentration (recall figure~\ref{fig:phiv_1wc}), directly linked to turbulence modulation via the two-way coupling stress term.

\section{Conclusions}
\label{sec:conclusion}

We have investigated the importance of lift forces in the dynamics of particle-laden turbulent channel flow. Three relevant shear-induced lift force models are analysed using PP-DNS for varying particle size and fixed Stokes number. Our results confirm that shear-induced lift forces strongly influence particle dynamics. We have analysed the main mechanisms for near-wall particle accumulation under lift forces: lift-induced migration, turbophoresis, and biased sampling. Very close to the wall, where the particle-to-fluid slip velocity is positive and particle fluctuations are vanishing, lift forces and turbophoresis cooperate in driving particles toward the wall. Particle--wall interactions and biased sampling balance this effect. Further away from the wall, lift forces and turbophoresis compete, with the former having a wall-repelling effect. Concerning biased sampling, we observe that, for larger particle sizes ($D^+ \geq 1$), the lift-free ascending particles prefer to sample ejections, while the descending ones tend to sample sweeps. Conversely, particles under lift forces tend to sample ejection regions while experiencing a marginal occurrence of sweep events, regardless of the employed lift model or wall-normal particle velocity sign. This is again a consequence of the competing wall-repelling effect of lift force. For the smallest particles ($D^+=0.1$), all cases exhibit over-sampling in ejection and inward motions of low-speed fluid regions (i.e., Q3 events). Overall, the sampling percentage of the ejections events only slightly surpasses that of Q3, suggesting that biased sampling plays a minor direct role in near-wall particle transport. However, the biased sampling of low-speed regions exposes particles for long times to persistent lift forces, which results in significantly reduced particle accumulation, even for small particle sizes.

These different dynamics have major consequences for turbulence modulation due to a modification of the near-wall values of (1) the local particle concentration, and (2) the correlated streamwise-wall-normal particle velocity fluctuations. While the latter is only modified for sufficiently large particle size and therefore can be seen as a finite-size effect, the former is still non-negligible for relatively small particle sizes, and can still result in small but visible changes in overall drag. Indeed, correlated particle velocity fluctuations show little sensitivity to lift forces for sufficiently small values of inner-scaled particle diameter ($D^+\lesssim 0.1$). Conversely, for small particle sizes, the modification of near-wall particle accumulation due to lift showed little sensitivity to the flavour of shear-induced lift model. Hence, to accurately account for lift-induced turbulence modulation with $D^+=O(1)$ particle sizes, reliable shear-induced lift models are needed that correctly reproduce the particle turbulent momentum flux. This could be seen as a necessary correction for finite-size effects, in the same spirit as the Fax\'en correction in the Maxey--Riley--Gatignol equation.

While such correction is important for reliable particle-modelled simulations of turbulent wall flows, our analysis has shown that current lift force models are bound to lose their validity in the near-wall region for small inertial particles. The reason is that the well-known tendency of particles to over-sample low-speed regions in the buffer layer, along with their tendency to flow faster than the fluid in the viscous sublayer, results in a region near the wall of vanishing particle-to-fluid slip velocity. Naturally, this issue will be even more pronounced at lower Stokes numbers, where the range of very low slip velocities in regions of high shear is wider. Unfortunately, current shear-induced lift force models are only available for non-dimensional shear rates $Sr$ of order unity, which robustly tends to infinity within the buffer layer. Hence, the condition of $Sr\leq1$ required by these models is seriously violated in wall turbulence with small inertial particles.


\backsection[Acknowledgements]{The Cray XC40 Shaheen II at KAUST was used for all simulations reported. PC thanks F.~Picano from Padova University for insightful discussions.}

\backsection[Funding]{This research was partially supported by the KAUST Office of Sponsored Research (OSR) under award no. OSR-2019-CCF-3666 and under baseline research funds of MP.}

\backsection[Declaration of interests]{The authors report no conflict of interest.}


\backsection[Author ORCIDs]{
\noindent Wei Gao,  https://orcid.org/0000-0001-7313-0058;
\noindent Pengyu Shi, https://orcid.org/0000-0001-6402-4720;
\noindent Matteo Parsani, https://orcid.org/0000-0001-7300-1280;
\noindent Pedro Costa, https://orcid.org/0000-0001-7010-1040.}


\bibliographystyle{jfm}
\bibliography{lift}

\end{document}